\def\uR{\boldsymbol{\hat{\rho}}}
\def\uT{\boldsymbol{\hat{\sigma}}}
\def\uN{\boldsymbol{\hat{\nu}}}

\def\nk{n_{\rm b}}

\def\Pb{P_{\rm b}}

\def\rfr#1{Equation~(\ref{#1})}
\def\rfrs#1#2{Equations~(\ref{#1})~to~(\ref{#2})}

\def\dert#1#2{\frac{{{\textrm{d}}}{#1}}{{{\textrm{d}}}{#2}}}

\def\eqi{\begin{equation}}
\def\eqf{\end{equation}}
\def\eqia{\begin{eqnarray}}
\def\eqfa{\end{eqnarray}}

\def\rp#1#2{{#1\over#2}}
\def\lb#1{\label{#1}}

\def\bds#1{\boldsymbol{#1}}

\def\ton#1{\left(#1\right)}
\def\qua#1{\left[#1\right]}
\def\grf#1{\left\{#1\right\}}

\documentclass[onecolumn]{aastex}

\usepackage{hyperref}
\usepackage{float}
\usepackage{amsmath,textgreek,w-greek,wasysym}
\usepackage{amsthm}
\usepackage{amscd,lineno}
\usepackage{amssymb,dsfont}
\usepackage{graphicx,epsfig}
\usepackage{txfonts}
\usepackage{xr-hyper}

\RequirePackage{color}

\newcommand{\emaila}{lorenzo.iorio@libero.it}

\allowdisplaybreaks[1]

\begin{document}

\title{Perspectives on constraining a cosmological constant-type parameter with pulsar timing in the Galactic Center}

\shortauthors{L. Iorio}

\author{Lorenzo Iorio\altaffilmark{1} }
\affil{Ministero dell'Istruzione, dell'Universit\`{a} e della Ricerca
(M.I.U.R.)-Istruzione
\\ Permanent address for correspondence: Viale Unit\`{a} di Italia 68, 70125, Bari (BA),
Italy}

\email{\emaila}

\begin{abstract}
Independent tests aiming to constrain the value of the cosmological constant $\Lambda$ are usually difficult because of its extreme smallness $\left(\Lambda \simeq 1\times 10^{-52}~\textrm{m}^{-2}\textcolor{black}{,~\textrm{or}~2.89\times 10^{-122}~\textrm{in~Planck~units}}\right)$. Bounds on it from Solar System orbital motions determined with spacecraft tracking are currently at the $\simeq 10^{-43}-10^{-44}~\textrm{m}^{-2}~\textcolor{black}{\left(5-1\times 10^{-113}~\textrm{in~Planck~units}\right)}$ level, but they may turn out to be somewhat optimistic since $\Lambda$ has not yet been explicitly modeled in the planetary data reductions. Accurate $\left(\sigma_{\tau_\textrm{p}}\simeq 1-10~\upmu\textrm{s}\right)$ timing of expected pulsars orbiting the Black Hole at the Galactic Center, preferably along highly eccentric and wide orbits,  might, at least in principle, improve the planetary constraints by several orders of magnitude.
By looking at the average time shift per orbit $\overline{\Delta\delta\tau}^\Lambda_\textrm{p}$, a S2-like orbital configuration with $e=0.8839,~\Pb=16~\textrm{yr}$ would allow to obtain \textcolor{black}{preliminarily} an upper bound \textcolor{black}{of the order of} $\left|\Lambda\right|\lesssim 9\times 10^{-47}~\textrm{m}^{-2}~\textcolor{black}{\left(\lesssim 2\times 10^{-116}~\textrm{in~Planck~units}\right)}$ \textcolor{black}{if only $\sigma_{\tau_\textrm{p}}$ were to be considered}.
Our results can be easily extended to modified models of gravity using $\Lambda-$type parameters.
\end{abstract}

keywords{
gravitation--pulsars: general--stars: black holes
}

%
\section{Introduction}\lb{intro}
The cosmological constant (CC) $\Lambda$ \citep{1989RvMP...61....1W,1992ARA&A..30..499C,2001LRR.....4....1C,2003RvMP...75..559P,2003PhR...380..235P,2004sgig.book.....C,Davis:2010,2017arXiv171106890O} is the easiest way to explain certain large-scale features of the universe like the acceleration of its expansion  \citep{1998AJ....116.1009R,1999ApJ...517..565P} and the growth of fluctuations by gravity \citep{2008PhRvD..77b3504N} within General Relativity (GR) assumed as a fundamental ingredient of the standard $\Lambda\textrm{CDM}$ model \citep{2015Sci...347.1100S}; for a recent overview of the status and future challenges of the Einsteinian theory of gravitation, see, e.g., \citet{2016Univ....2...23D}. \textcolor{black}{Interestingly, the CC was considered before Einstein for possible modification of the Poisson equation in the framework of the Newtonian gravity \citep{1895AN....137..129S}.}
The CC can be expressed in terms of the Hubble parameter $H_0$ and the ratio $\Omega_\Lambda$ between the density due to the cosmological constant itself $\varrho_\Lambda = \ton{1/8\uppi} c^2 \Lambda G^{-1}$ and the critical density $\varrho_\textrm{crit} = \ton{3/8\uppi} H_0^2 G^{-1}$ as $\Lambda = 3 H_0^2 \Omega_\Lambda c^{-2}$, where \citep{2016A&A...594A..13P} $H_0= 67.74\pm 0.46~\textrm{km~s}^{-1}~\textrm{Mpc}^{-1},~\Omega_\Lambda= 0.6911\pm 0.0062$. As such, its most recent value inferable from the  measurements of the Cosmic Microwave Background (CMB) power spectra by the satellite Planck reads
\eqi
\Lambda = (1.11\pm 0.02)\times10^{-52}~\textrm{m}^{-2}\lb{Lambda}.
\eqf
\textcolor{black}{
In order to relate it to possible symmetry breaking in gravity \citep{2011PhLB..702..187M}, the CC is sometimes written as a very tiny dimensionless parameter essentially by multiplying it by the square of the Planck length $\ell_\textrm{P}=\sqrt{\hbar G c^{-3}}=1.61\times 10^{-35}~\textrm{m}$. Thus, one gets, in Planck units,
\eqi
\Lambda = 2.89\times 10^{-122}.
\eqf
}
A CC-type parameterization occurs also in several classes of long range modified models of gravity aiming to explain in a unified way seemingly distinct features of the cosmic dynamics like inflation, late-time acceleration and even dark matter \citep{2006hep.th....1213N,2007JPhA...40.6725N,2010PhRvD..82b3519D,2010LRR....13....3D,
2011PThPS.190..155N,2011PhR...509..167C,2012PhR...513....1C,2012AnP...524..545C,2015Univ....1..199C,2015Univ....1..123D,
2015IJMPD..2441002C,2016RPPh...79j6901C}.

Ever since the time of Einstein, who, on the backdrop of what is mathematically feasible with the Poisson equation, included $\Lambda$ in his GR field equations to obtain a non-expanding, static cosmological model \citep{1917SPAW.......142E}, the introduction of the CC has always been justified from an observational/experimental point of view by arguing that it would not be in contrast with any observed effects in local systems like, e.g., orbital motions in gravitationally bound binary systems because of its extreme smallness. As a consequence, there are not yet independent, non-cosmological tests of the CC itself for which only relatively loose constraints from planetary motions of the Solar System exist in the literature. So far, most of the investigations on the consequences  of the CC in local binary systems have focused on the anomalous pericenter precession induced by $\Lambda$ \citep{Islam83,Car98,2001rsgc.book.....R,Kerr03,Kran03,Iorio06,Jetz06,Kagra06,Ser06,Adk07a,
Adk07b,2007JCAP...01..010R,Ser07,Iorio08,2008PhRvD..77j7502C,2012MNRAS.427.1555I,2013IJTP...52.1408A,
2013MNRAS.433.3584X,2015JCAP...08..021I,2015arXiv151104829O} on the basis of a Hooke-type perturbing potential \citep{2001rsgc.book.....R,Kerr03}
\eqi
U_\Lambda= -\rp{1}{6}\Lambda c^2 r^2\lb{Ulam}
\eqf
arising in the framework of the Schwarzschild-de Sitter spacetime \citep{1918AnP...361..401K,1999PhRvD..60d4006S,2001rsgc.book.....R}.
Equation~\ref{Ulam} yields the radial extra-acceleration \citep{2001rsgc.book.....R,Kerr03}
\eqi
{\bds A}_\Lambda = \rp{1}{3}\Lambda c^2\mathbf{r}.\lb{Accel}
\eqf
The latest upper limits on the absolute value of $\Lambda$, inferred within the framework of $f\ton{T}$ gravity from the anomalous perihelion precessions of some of the planets of the Solar System tightly constrained with the INPOP10a ephemerides \citep{2011CeMDA.111..363F}, are of the order of \citep{2013MNRAS.433.3584X}
\eqi
\left|\Lambda\right| \lesssim 2\times 10^{-43}~\textrm{m}^{-2}\lb{bounda},
\eqf
\textcolor{black}{
corresponding to
\eqi
\left|\Lambda\right| \lesssim 5\times 10^{-113}
\eqf
in Planck units.
}
The Earth-Saturn range residuals constructed from the telemetry of the Cassini spacecraft \citep{2014PhRvD..89j2002H} yielded an upper limit of the order of \citep{2015JCAP...08..021I}
\eqi
\left|\Lambda\right| \lesssim 5\times 10^{-44}~\textrm{m}^{-2}\lb{boundb},
\eqf
\textcolor{black}{i.e.
\eqi
\left|\Lambda\right| \lesssim 1\times 10^{-113}
\eqf
in Planck units.
}
\citet{2016PDU....13..111I} suggested that a challenging analysis of the telemetry of the New Horizons spacecraft might improve the limit of \rfr{boundb} by about one order of magnitude.
On the other hand, the bounds of \rfrs{bounda}{boundb} may be somehow optimistic since they were inferred without explicitly modeling \rfr{Accel} in the dynamical force models of the ephemerides. As such, its signature may have been removed from the post-fit residuals to a certain extent, being partially absorbed in the estimation of, e.g., the planets' initial state vectors.
\textcolor{black}{
Such a possibility was investigated  by simulating observations of major bodies of Solar System in the case of some modified models of gravity \citep{2012CQGra..29w5027H}.
}
Thus, more realistic constraints might yield larger values for the allowed upper bound on $\Lambda$.

In this paper, we will show that the future, long waited discovery of pulsars revolving around the \textcolor{black}{putative} Supermassive Black Hole (SMBH) in the Galactic Center (GC) at Sgr A$^\ast$ \citep{2004ApJ...615..253P,2014ApJ...784..106Z,2014MNRAS.440L..86C,2017MNRAS.471..730R} along sufficiently wide and eccentric orbits and their timing accurate to the $\sigma_{\tau_\textrm{p}}\simeq 1-10~\upmu\textrm{s}$ level \citep{2016ApJ...818..121P,2017IJMPD..2630001G}, might allow, in principle, to substantially improve the planetary bounds of \rfrs{bounda}{boundb} by several orders of magnitude, getting, perhaps, closer to the level of \rfr{Lambda} itself under certain fortunate conditions.
\textcolor{black}{
The possibility that travelling gravitational waves can be used in a foreseeable future for local measurements of the CC through their impact on Pulsar Timing Arrays (PTA) is discussed in \citet{2014AIPC.1606...86E}.
}
In Section~\ref{calcolo} we will analytically work out the perturbation $\Delta\delta\tau^\Lambda_\textrm{p}$ induced by $\Lambda$ on the  the pulsar's timing periodic variation $\delta\tau_\textrm{p}$ due to its orbital motion around the SMBH; we will follow the approach put forth in \citet{2017EPJC...77..439I} applying it to \rfr{Accel}. We will neglect the time shifts due to the CC on the propagation of the electromagnetic waves \citep{2008A&A...484..103S}. Despite it can be shown that, for certain values of the initial conditions, an extremely wide  orbital configuration like, say, that of the actually existing star S85 may yield values of the instantaneous changes $\Delta\delta\tau^\Lambda_\textrm{p}\ton{t}$ as large as just $\simeq 1-10~\upmu\textrm{s}$, caution is in order because of, e.g., the very likely systematic bias induced on such an extended orbit by the poorly known mass background in the GC \citep{2011PhRvD..84d4024M,2011CQGra..28v5029S,2014MNRAS.444.3780A,2017ApJ...834..198Z}. Moreover, also the accurate knowledge of the SMBH physical parameters like mass, angular momentum and quadrupole moment would be of crucial importance because of the competing pN orbital timing signatures $\Delta\delta\tau^\textrm{pN}_\textrm{p}$ which would superimpose to the CC effect. Finally, also the orbital parameters of the pulsar should be determined over a relatively short time interval $\Delta T$ with respect to its extremely long orbital period $P_\textrm{b}$. If, instead, a closer pulsar is considered, it makes sense to look at its net orbital time shift per orbit $\overline{\Delta\delta\tau}_\textrm{p}^\Lambda$. \citet{2017arXiv170908341Z} recently investigated the possibility of constraining the SMBH's spin with such kind of rapidly orbiting pulsars. See also \citet{2017arXiv171200265D}. In Section~\ref{misura} it will be shown that a S2-type orbital geometry, summarized in Table~\ref{S2}, would allow, in principle, to improve the planetary bounds of \rfrs{bounda}{boundb} by about $3-4$ orders of magnitude. A strategy to overcome the potentially serious bias posed by the competing post-Newtonian (pN) orbital time delays driven by the SMBHS's mass, spin and quadrupole moment will be discussed as well. In Section~\ref{fine}, we summarize our findings and offer our conclusions.
\section{Calculating the perturbation of the orbital component of the time shift due to the cosmological constant}\lb{calcolo}
Here, the analytical method devised in \citet{2017EPJC...77..439I}, relying upon \citet{1993CeMDA..55..209C}, will be applied to the perturbing acceleration of \rfr{Accel} with some technical modifications. Indeed, since, in this case, the use of the eccentric anomaly $E$ as fast variable of integration instead of the true anomaly $f$ turns out to be computationally more convenient, Equations~(30)~to~(31) of \citet{1993CeMDA..55..209C}, giving the radial and transverse components of the perturbation $\Delta\mathbf{r}$ of the position vector $\mathbf{r}$ and used in \citet{2017EPJC...77..439I} as Equations~(3)~to~(4), have to be replaced with Equations~(36)~to~(37) of \citet{1993CeMDA..55..209C}, i.e.
\begin{align}
\Delta\textrm{r}_\rho\ton{E} \lb{Dr}&=\rp{r\ton{E}}{a}~\Delta a\ton{E} -\rp{r\ton{E}\ton{e + \cos f}}{1-e^2}~\Delta e\ton{E} + \rp{r\ton{E}e\sin f}{\sqrt{1-e^2}}~\Delta E\ton{E}, \\ \nonumber \\
\Delta\textrm{r}_\sigma\ton{E} \lb{Dtau}&= \rp{r\ton{E}\sin f }{1-e^2}~\Delta e\ton{E} + a \sqrt{1-e^2}~\Delta E\ton{E} + r\ton{E}\qua{\cos I~\Delta\Omega\ton{E}+\Delta\omega\ton{E}}.
\end{align}
Equation~(32) of \citet{1993CeMDA..55..209C}, giving the out-of-plane component $\Delta\textrm{r}_\nu$ of the perturbation $\Delta\mathbf{r}$ of the position vector $\mathbf{r}$ and used in \citet{2017EPJC...77..439I} as Equation~(5), remains unchanged.
Thus, the perturbation of the $z$ component of the pulsar's position vector $\mathbf{r}$ reads
\begin{align}
\Delta\textrm{r}_z \nonumber &= \rp{r\ton{E}}{a}\sin I\sin u~\Delta a\ton{E}-\rp{r\ton{E}\sin I\ton{\sin\omega+e\sin u}}{1-e^2}~\Delta e\ton{E}+r\ton{E}\cos I\sin u~\Delta I\ton{E}+ \\ \nonumber\\
&+ r\ton{E}\sin I\cos u~\Delta\omega\ton{E} + \rp{\sin I\qua{a\ton{1-e^2}\cos u +er\ton{E}\sin f\sin u  }}{\sqrt{1-e^2}}~\Delta E\ton{E}\lb{Dz}.
\end{align}
From \citet{2017EPJC...77..439I}, it is $\Delta\delta\tau_\textrm{p} = \Delta\textrm{r}_z c^{-1}$ in a coordinate system whose reference $z$ axis points towards the observer perpendicularly to the plane of the sky spanned by the reference $\grf{x,~y}$ plane.
In \rfrs{Dr}{Dz}, the instantaneous shift $\Delta E\ton{E}$ of the eccentric anomaly can be expressed, in turn, in terms of the perturbations $\Delta\mathcal{M}\ton{E},~\Delta e\ton{E}$ of the mean anomaly and the eccentricity, respectively, according to Equation~(A.5) of \citet{1993CeMDA..55..209C}, i.e.
\eqi
\Delta E\ton{E} = \rp{a}{r\ton{E}}\qua{\Delta\mathcal{M}\ton{E}+\sin E~\Delta e\ton{E}}.\lb{casot}
\eqf
The instantaneous shifts  of the osculating orbital elements are to be computed in terms of $E$ as
\eqi
\Delta\kappa\ton{E} =\int_{E_0}^E\dert{\kappa}{t}\dert{t}{E^{'}}dE^{'},~\kappa=a,~e,~I,~\Omega,~\omega
\eqf
with the aid of the standard formulas of celestial mechanics
\begin{align}
\sin f & = \rp{\sqrt{1-e^2}~\sin E}{1 - e\cos E}\lb{sinf}, \\ \nonumber\\
\cos f & = \rp{\cos E -e}{1 - e\cos E}, \\ \nonumber \\
r\ton{E} & = a\ton{1- e\cos E}\lb{rE}, \\ \nonumber \\
\dert{t}{E} & = \rp{1 - e\cos E}{\nk}
\end{align}
applied to the usual Gauss equations for the variation of the elements yielding $d\kappa/dt$.
The calculation of the perturbation $\Delta\mathcal{M}\ton{E}$ of the mean anomaly has to be performed as shown in \citet{2017EPJC...77..439I}, whose
Equations~(20)~to~(21) are to be calculated with $E$.
The CC-induced instantaneous perturbations of the osculating orbital elements turn out to be
\begin{align}
\Delta a\ton{E} & = \rp{c^2\Lambda a e\ton{\cos E-\cos E_0}\qua{-2 +e\ton{\cos E+\cos E_0}} }{3 \nk^2 }\lb{sma}, \\ \nonumber \\
\Delta e\ton{E} & = \rp{c^2\Lambda\ton{1-e^2}\ton{\cos E-\cos E_0}\qua{-2 +e\ton{\cos E+\cos E_0}} }{6 \nk^2 }\lb{ecce}, \\ \nonumber \\
\Delta I\ton{E} & = 0, \\ \nonumber \\
\Delta \Omega\ton{E} & = 0, \\ \nonumber \\
\Delta \omega\ton{E} \nonumber & = \rp{c^2\Lambda\sqrt{1-e^2}}{12e\nk^2}\qua{4 \ton{1 + e^2} \sin E_0 - e \ton{6 E_0 - 6 E + \sin 2 E_0} -\right.\\ \nonumber \\
&\left. -  4 \ton{1 + e^2} \sin E + e \sin 2 E}\lb{omega}, \\ \nonumber \\
\Delta\mathcal{M}\ton{E}\nonumber & = \rp{c^2\Lambda}{72 e\nk^2 }\grf{12 e \ton{7 + 6 e^2} \ton{E_0 - E} - 4 \ton{6 + 54 e^2 + 7 e^4} \sin E_0 +
6 e \sin 2 E_0 + \right.\\ \nonumber \\
\nonumber &\left. + 3 \ton{8 + 72 e^2 + 7 e^4} \sin E + 2 e^3  \qua{9\ton{E-E_0} + e\ton{2\sin E_0 - 9\sin E}}\cos 2 E_0 +\right.\\ \nonumber \\
\nonumber &+\left. 6 e^2  \qua{7 e \sin E_0 + 12 \ton{E_0 - E + e \sin E}}\cos E_0 -\right.\\ \nonumber \\
&-\left.  3 e \ton{2 + 19 e^2} \sin 2 E + 7 e^4 \sin 3 E}.\lb{manom}
\end{align}
By inserting \rfr{ecce} and \rfr{manom} in \rfr{casot}, it is possible to explicitly infer the instantaneous perturbation of the eccentric anomaly
\begin{align}
\Delta E\ton{E} \nonumber & = -\rp{c^2\Lambda}{72 e\nk^2\ton{1-e\cos E} }\grf{12\qua{e\ton{7 + 6e^2}\ton{E-E_0}  +2\ton{\sin E_0 - \sin E}  } + \right.\\ \nonumber\\
\nonumber &\left. + e\qua{ 6e\ton{36+5e^2}\sin E_0  -3\ton{2+7e^2}\sin 2 E_0  -2e^3\sin 3E_0  -3e\ton{71+8e^2}\sin E +\right.\right.\\ \nonumber \\
\nonumber &\left.\left.  +\ton{3 e E_0 - 3 e E + \sin E + 2 e^2 \sin E}\ton{6e\cos2E_0-24\cos E_0}  +\right.\right.\\ \nonumber \\
&+\left.\left. 9\ton{2 + 5e^2}\sin 2 E -e\ton{3+4e^2}\sin 3 E  }}.\lb{eccanom}
\end{align}
By inserting \rfrs{sma}{omega} and \rfr{eccanom} in \rfr{Dz} and using \rfrs{sinf}{rE} allows one to obtain the instantaneous perturbation $\Delta\delta\tau_\textrm{p}^\Lambda\ton{E}$ of the orbital time shift of the pulsar p due to $\Lambda$. It is
\eqi
\Delta\delta\tau_\textrm{p}^\Lambda\ton{E} = \rp{c\Lambda a\sin I}{72\nk^2}\mathcal{L}\ton{E;~E_0,~e,~\omega}\lb{DTmega},
\eqf where $\mathcal{L}\ton{E;~E_0,~e,~\omega}$ is a function of  $E$ and  the parameters $E_0,~e,~\omega$ definitely too cumbersome to be explicitly displayed. Thus, we show only the leading term of \rfr{DTmega};
\begin{align}
\Delta\delta\tau_\textrm{p}^\Lambda\ton{E} \nonumber &\simeq \rp{c\Lambda a\sin I}{6\nk^2}\qua{4 \ton{E_0 - E} \cos\ton{E + \omega} - \sin\ton{E_0 - 2 E - \omega} - \right.\\ \nonumber \\
&\left. - 3 \sin\ton{E_0 + \omega} + 2 \sin\ton{E + \omega}} +\mathcal{O}\ton{e^k},~k\geq 1.\lb{Dtau}
\end{align}
It is important to note from \rfr{DTmega} that $\Delta\delta\tau_\textrm{p}^\Lambda$ is proportional to the fourth power of the semimajor axis $a$, which characterizes the size of the pulsar's orbit, and is inversely proportional to the mass of the SMBH.

The net shift per orbit can be calculated from \rfr{DTmega} with $E\rightarrow E_0+2\uppi$: it turns out to be
\begin{align}
\overline{\Delta\delta\tau}_\textrm{p}^\Lambda \nonumber &= -\rp{\uppi c\Lambda a\sin I}{12\nk^2}\rp{1}{\ton{1-e\cos E_0}}\grf{\sqrt{1 - e^2} \qua{\ton{16 + 9 e^2} \cos E_0 + \right.\right.\\ \nonumber\\
\nonumber &\left.\left. + 3 e \ton{10 + 6 \cos 2E_0 - e \cos 3 E_0}} \cos\omega - 16 \sin E_0 \sin\omega + \right.\\ \nonumber \\
&\left. + 6 e \qua{2 \ton{-3 + e^2} \cos E_0 + e \ton{-6 + \cos 2E_0}} \sin E_0 \sin\omega }.\lb{DTave}
\end{align}
It can be noted that also \rfr{DTave} depends on the initial conditions through $E_0$.
It is also important to stress that both \rfr{DTmega} and \rfr{DTave} were worked out without any a priori simplifying approximations about the pulsar's orbital configuration; they hold for all values of $e$. It is a key feature in view of the highly eccentric orbits revealed so far in the GC.
\section{The opportunity offered by hypotetical pulsars in the Galactic Center}\lb{misura}
\textcolor{black}{Let us, now, move to the compact object located in Sgr A$^\ast$. For an interesting multidisciplinary discussion about the possibility that it is, actually, a SMBH or something else, see the recent overview in \citet{2017FoPh...47..553E}. However, our results will be unaffected by the alternative possibilities discussed there since their spacetimes are undistinguishable from that of a SMBH for the pulsars' orbital motions of interest here.}

\begin{figure*}
\centerline{
\vbox{
\begin{tabular}{c}
\epsfbox{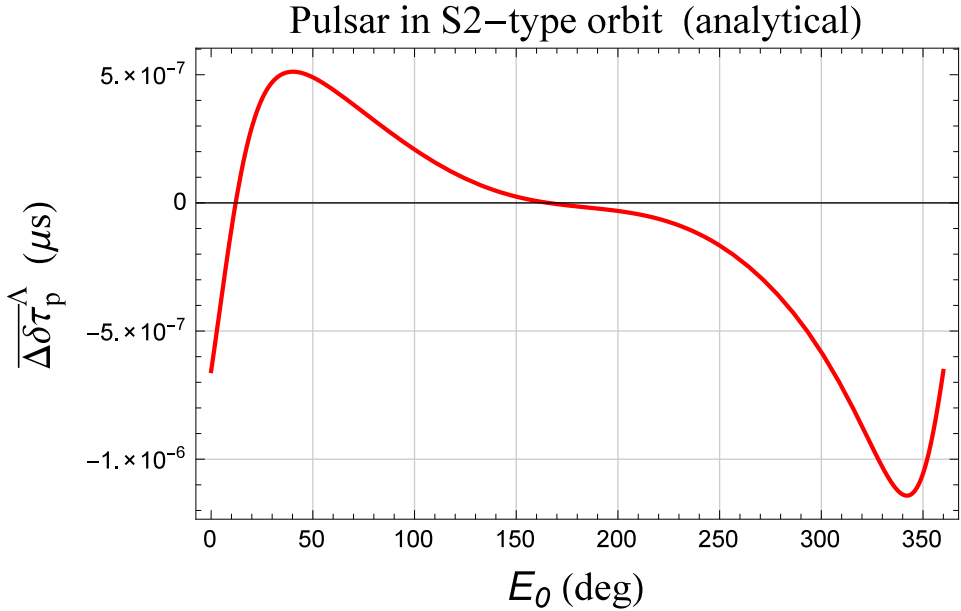}\\
\end{tabular}
}
}
\caption{Average orbital time shift per orbit $\overline{\Delta\delta\tau}_\textrm{p}^\Lambda$, in $\upmu\textrm{s}$,  of a hypothetical pulsar in Sgr A$^\ast$ obtained analytically from \rfr{DTave} along with the value of \rfr{Lambda} for $\Lambda$ as a function of the initial phase $E_0$. The orbital configuration of the S2 star, quoted in Table \ref{S2}, was adopted. It can be noted that $\overline{\Delta\delta\tau}_\textrm{p}^\Lambda$ vanishes for two given values of $E_0$; the largest absolute value occurs for $E_0=342.08~\textrm{deg}$. By assuming a pulsar timing accuracy of $\sigma_{\tau_\textrm{p}}=1~\upmu\textrm{s}$, it translates to an upper bound on $\Lambda$ of the order of $\left|\Lambda\right|\leq 9\times 10^{-47}~\textrm{m}^{-2}~\left(\lesssim 2\times 10^{-116}~\textrm{in~Planck~units}\right)$.}\label{Fig1}
\end{figure*}

In order to explore the opportunity offered by our results to effectively constrain the CC with pulsar timing in the GC, let us consider a putative pulsar whose orbital period $\Pb$ is short enough to allow to monitor at least one full revolution during a timing campaign. In this case, by suitably choosing the initial orbital phase $E_0$, it would be possible to profitably use \rfr{DTave} in order to maximize it; indeed, in principle, \rfr{DTave} can even vanish. To this aim, for the sake of concreteness, let us assume a S2-type orbital configuration characterized by $\Pb=16~\textrm{yr},~e=0.8839$ \citep{2017ApJ...837...30G}. It turns out that the maximum of the absolute value of \rfr{DTave} occurs for $E_0=342.08~\textrm{deg}$, which corresponds to almost an orbital period after the time of periastron passage, yielding an upper bound on the CC as little as
\eqi
\left|\Lambda\right|\lesssim 9\times 10^{-47}~\textrm{m}^{-2}~\left(\lesssim 2\times 10^{-116}~\textrm{in~Planck~units}\right)\lb{boundS2}
\eqf
for a timing accuracy of $\sigma_{\tau_\textrm{p}}\simeq 1~\textrm{\upmu s}$. It should be noted that \rfr{boundS2} is $3-4$ orders magnitude better than the (likely optimistic) planetary bounds of \rfrs{bounda}{boundb}. Fig.~\ref{Fig1} depicts the plot of \rfr{DTave} as a function of $E_0$. If we modify some of the parameters of the pulsar's orbital configuration by adopting, say, $\Pb=30~\textrm{yr},~e=0.987,~I=90~\textrm{deg}$, it is possible to improve the bound on the CC to the level
\eqi
\left|\Lambda\right|\lesssim 4\times 10^{-48}~\textrm{m}^{-2}~\textcolor{black}{\left(\lesssim 1\times 10^{-117}~\textrm{in~Planck~units}\right)}\lb{boundX}
\eqf
for $E_0=354.04~\textrm{deg}$.
\textcolor{black}{About the figures in \rfrs{boundS2}{boundX}, inferred by considering only $\sigma_{\tau_\textrm{p}}$ as source of observational error, it must be stressed that they should be regarded with caution as preliminary and just indicative of the potential of the approach proposed. If not explicitly modeled and simultaneously estimated in actual pulsar timing data reductions, the CC-induced signature may be partially removed from the resulting residual. As such,  the resulting bounds may be weaker than those in \rfrs{boundS2}{boundX}. Further dedicated analyses should be made by simulating observations and fitting a full orbital model to them in order to assess how good the input values are recovered. }
A possible source of systematic uncertainty is represented by the mismodelled part of the competing averaged orbital time shifts induced by the standard post-Newtonian (pN) effects due to the current experimental errors in the SMBH's parameters entering their formulas. For example, according to Equation~(35) of \citet{2017EPJC...77..439I}, the amplitude of the 1pN gravitoelectric average time shift $\overline{\Delta\delta\tau}_\textrm{p}^\textrm{GE}$ is proportional to $\mu_\bullet c^{-3}=22~\textrm{s}$, while the mass of the SMBH is currently known at a $\simeq 7\%$ level of accuracy \citep{2017ApJ...837...30G}. Analogous considerations hold for the Lense-Thirring (Equation~(51) of \citet{2017EPJC...77..439I}) and quadrupole (Equation~(83) of \citet{2017EPJC...77..439I}) average shifts. In principle, such an issue could be circumvented if $N$ pulsars $j$ with different orbital configurations will be discovered. Indeed, in this case, it could be possible to write down for each of them an analytical expression
\eqi
\overline{\Delta\delta\tau}^\textrm{exp}_j = \overline{\Delta\delta\tau}^\textrm{GE}_j+\overline{\Delta\delta\tau}^\textrm{LT}_j+
\overline{\Delta\delta\tau}^{Q_2}_j+\overline{\Delta\delta\tau}^\Lambda_j,~j=1,2,\ldots N\lb{linear}
\eqf
for their measured average orbital time shift $\overline{\Delta\delta\tau}^\textrm{exp}_j$ as a sum of the pN terms plus the CC one by treating $\mu_\bullet,~S_\bullet,~Q_2^\bullet,~\Lambda$, which enter each term of \rfr{linear} as multiplicative scaling parameters, as unknowns of the resulting linear system of algebraic equations. Solving for them, it would be possible to obtain, among other things,  an expression for $\Lambda$ independent, by construction, of the mismodeled SMBH's physical parameters. Such an approach could be extended also to other dynamical effects impacting the pulsar's average orbital time shift like, e.g., third-body perturbations.

\textcolor{black}{
Recently, the upper bound
\eqi
\left|\dot\omega_\textrm{S2}\right|\lesssim 1.6\times 10^{-3}~\textrm{yr}^{-1}=9.2~\textrm{deg~cty}^{-1}\lb{hees}
\eqf
on the periastron precession of the real star S2 was inferred in \citet{2017PhRvL.118u1101H}. By combining \rfr{hees} with the well known analytical expression for the $\Lambda-$induced  pericenter precession (see the references cited in Section~\ref{intro})
\eqi
\dot\omega_\Lambda=\rp{1}{2}\ton{\rp{\Lambda c^2}{\nk}}\sqrt{1-e^2},
\eqf
it is possible to infer a tentative upper limit on the CC of the order of
\eqi
\left|\Lambda\right|\lesssim 3\times 10^{-35}~\textrm{m}^2~\left(\lesssim 8\times 10^{-105}~\textrm{in~Planck~units}\right).
\eqf
}
\textcolor{black}{
For much more distant pulsars, major sources of systematic uncertainty would be given by the still poorly mass background and the difficulty of effectively constraining the  parameters of extremely wide orbits \citep{2014A&A...563A.126L} and of the Black Hole itself over a relatively short observational time interval $\Delta T$ with respect to the expected extremely long orbital period $P_\textrm{b}$ of the neutron star.
}
%
%
%
\section{Summary and conclusions}\lb{fine}
In this paper, we analytically calculated the perturbation $\Delta\delta\tau_\textrm{p}^\Lambda$ induced by the CC $\Lambda$ on the orbital part of the time variation $\delta\tau_\textrm{p}$ of a hypothetical pulsar p orbiting the SMBH in Sgr A$^\ast$. We did not restrict to any particular orbital configuration, and our results are, thus, exact with respect to the eccentricity $e$; it is an important feature since most of the main sequence stars discovered so far in the GC move along highly eccentric orbits. We obtained both the instantaneous change $\Delta\delta\tau_\textrm{p}^\Lambda\ton{E}$ and the net shift per orbit $\overline{\Delta\delta\tau}_\textrm{p}^\Lambda$: they are proportional to $c\Lambda a^4\sin I\mu_\bullet^{-1}$. A distinctive feature of both of them is their explicit dependence on the initial value $E_0$ of the orbital phase. Our results hold also for a wide class of long-range modified models of gravity generating an extra-potential quadratic in the distance $r$.

We applied our results to some putative scenarios by adopting, for the sake of definiteness, the orbital configurations of one actually existing main sequence star orbiting Sgr A$^\ast$. By considering a S2-type orbit with $\Pb = 16~\textrm{yr}$, it is meaningful to look at the averaged time shift $\overline{\Delta\delta\tau}_\textrm{p}^\Lambda$. It turns out that, for a careful choice of the initial orbital phase $E_0$, it would be possible, in principle, to infer an upper bound $\left|\Lambda\right|\lesssim 9\times 10^{-47}~\textrm{m}^{-2}$\textcolor{black}{, corresponding to $\lesssim 2\times 10^{-116}$ in Planck units,} by assuming a pulsar timing accuracy of $\sigma_{\tau_\textrm{p}}\simeq 1~\upmu\textrm{s}$. It would be $3-4$ orders of magnitude better than the current, likely optimistic, constraints from Solar System's planetary orbital motions.
\textcolor{black}{On the other hand, it should be stressed that the very same aforementioned bound on $\Lambda$, derived by accounting for only $\sigma_{\tau_\textrm{p}}$, may be optimistic in view of possible partial removal of the sought signature if not explicitly modeled and solved for in actual data reductions. As a suggestion for further dedicated investigations, simulating the observations and fitting a complete dynamical orbital model to them  would be needed in order to assess how accurately the input values can be recovered.}
The bias due to the errors in the physical parameters of the SMBH entering the competing pN net shifts per orbit could be eliminated by setting up  suitably designed linear combinations of the time delays measured for several pulsars. In the case of much more distant pulsars, using the orbital averaged time shift $\overline{\Delta\delta\tau}_\textrm{p}^\Lambda$ is unfeasible; only instantaneous values $\Delta\delta\tau_\textrm{p}^\Lambda\ton{E}$ could be, in principle, measured.
On the other hand, too wide and slow orbits may be impacted by the still poorly known mass background in the GC, and it would be difficult to effectively constrain the pulsar's orbital parameters over a relatively short time interval with respect to its extremely long orbital period.
\section*{Acknowledgements}
I would like to thank two attentive referee for their precious critical remarks.
\appendix
\section{Notations and definitions}\lb{appen}
Some basic notations and definitions used in the text are listed below \citep{1991ercm.book.....B,Nobilibook87,1989racm.book.....S,2003ASSL..293.....B}. In the case treated in this  paper, the unseen companion c of the pulsar p is the SMBH of mass $M_\bullet$, so that $m_\textrm{c}=M_\bullet\gg m_\textrm{p}$ and $a_\textrm{p}\simeq a$.
\begin{description}
\item[] $G:$ Newtonian constant of gravitation
\item[] $c:$ speed of light in vacuum
\textcolor{black}{
\item[] $\hbar:$ reduced Planck constant
\item[] $\ell_\textrm{P}\doteq\sqrt{\hbar G c^{-3}}:$ Planck length
}
\item[] $\Lambda:$ cosmological constant
\item[] $H_0:$ Hubble parameter
\item[] $\varrho_\textrm{crit}\doteq \ton{3/8\uppi}H_0^2 G^{-1}:$ critical density of the universe
\item[] $\varrho_\Lambda\doteq \ton{1/8\uppi}c^2\Lambda G^{-1}:$ density due to the cosmological constant
\item[] $\Omega_\Lambda\doteq\varrho_\Lambda\varrho_\textrm{crit}^{-1}:$ normalized energy density of the cosmological constant
\item[] $m_\textrm{p}$: mass of the pulsar p
\item[] $m_\textrm{c}$: mass of the invisible companion c
\item[] $m_\textrm{tot}\doteq m_\textrm{p} + m_\textrm{c}$: total mass of the binary
\item[] $\mu\doteq Gm_\textrm{tot}:$ gravitational parameter of the binary
\item[] $a:$  semimajor axis of the binary's relative orbit
\item[] $\nk \doteq \sqrt{\mu a^{-3}}:$   Keplerian mean motion
\item[] $\Pb = 2\uppi \nk^{-1}:$ Keplerian orbital period
\item[] $a_\textrm{p}=m_\textrm{c} m^{-1}_\textrm{tot} a:$ semimajor axis of the barycentric orbit of the pulsar p
\item[] $e:$  eccentricity
\item[] $I:$  inclination of the orbital plane
\item[] $\omega:$  argument of pericenter
\item[] $t_p:$ time of periastron passage
\item[] $t_0:$ reference epoch
\item[] $\mathcal{M}\doteq \nk\ton{t - t_p}:$ mean anomaly
\item[] $f:$  true anomaly
\item[] $E:$ eccentric anomaly
\item[] $u\doteq \omega + f:$  argument of latitude
\item[] $\mathbf{r}:$ relative position vector of the binary's orbit
\item[] $\textrm{r}_z:$ component of the position vector along the line of sight
\item[] $r:$ magnitude of the binary's relative position vector
\item[] $\uR:$ radial unit vector
\item[] $\uN:$ unit vector of the orbital angular momentum
\item[] $\uT\doteq\uN\bds\times\uR:$ transverse unit vector
\textcolor{black}{
\item[] $\textrm{r}_\rho:$ radial component of the relative position vector of the binary's orbit
\item[] $\textrm{r}_\nu:$ normal component of the relative position vector of the binary's orbit
\item[] $\textrm{r}_\sigma:$ transverse component of the relative position vector of the binary's orbit
}
\item[] $U_\Lambda:$ perturbing potential due to the cosmological constant
\item[] ${\bds A}_\Lambda:$ perturbing acceleration due to the cosmological constant
\item[] $\delta\tau_\textrm{p}=\textrm{r}_z c^{-1}:$ periodic variation of the time of arrivals of the pulses from the pulsar p due to its barycentric orbital motion
\end{description}
\section{Tables and Figures}
%
%
%
%
%
%
%
%
%
%
%
%
%
%
%
%
%
\begin{table*}
\caption{Relevant physical and orbital parameters of the S2 star and the SMBH at the GC along with their estimated uncertainties according to Table 3 of \citet{2017ApJ...837...30G}; they are referred to the epoch $2000.0$. $D_0$ is the distance to $\textrm{Sgr~A}^\ast$. The  linear size of the semimajor axis of S2 is $a=1044~\textrm{au}$. }
\label{S2}
\centering
\begin{tabular}{ll}
\noalign{\smallskip}
\hline
Estimated parameter & Value \\
\hline
$M_\bullet$ & $4.28\pm \left.0.10\right|_\textrm{stat}\pm \left.0.21\right|_\textrm{sys}\times 10^6~\textrm{M}_\odot$\\
$D_0$ & $8.32\pm\left.0.07\right|_\textrm{stat}\pm \left.0.14\right|_\textrm{sys}~\textrm{kpc}$\\
$\Pb$ & $16.00\pm 0.02~\textrm{yr}$\\
$a$ & $0.1255\pm 0.0009~\textrm{arcsec}$\\
$e$  & $0.8839\pm 0.0019$\\
$I$ & $134.18\pm 0.40~\textrm{deg}$\\
$\Omega$ & $226.94\pm 0.60~\textrm{deg}$\\
$\omega$ & $65.51\pm 0.57~\textrm{deg}$\\
$t_p$ & $2002.33\pm 0.01$~\textrm{calendar~year}\\
\hline
\end{tabular}
\end{table*}
%
%
%
%
%
%
%
%
%
%
\bibliography{PXbib,IorioFupeng,fTbib}{}

\begin{thebibliography}{79}
\expandafter\ifx\csname natexlab\endcsname\relax\def\natexlab#1{#1}\fi

\bibitem[{Adkins \& McDonnell(2007)}]{Adk07b}
Adkins G., McDonnell J., 2007, Phys. Rev. D, 75, 082001

\bibitem[{Adkins, McDonnell \& Fell(2007)Adkins, McDonnell, \& Fell}]{Adk07a}
Adkins G., McDonnell J., Fell R., 2007, Phys. Rev. D, 75, 064011

\bibitem[{{Ang{\'e}lil} \& {Saha}(2014)}]{2014MNRAS.444.3780A}
{Ang{\'e}lil} R., {Saha} P., 2014, MNRAS, 444, 3780

\bibitem[{{Arakida}(2013)}]{2013IJTP...52.1408A}
{Arakida} H., 2013, Int. J. Theor. Phys., 52, 1408

\bibitem[{{Bertotti}, {Farinella} \& {Vokrouhlick\'{y}}(2003){Bertotti},
  {Farinella}, \& {Vokrouhlick\'{y}}}]{2003ASSL..293.....B}
{Bertotti} B., {Farinella} P., {Vokrouhlick\'{y}} D., 2003, {Physics of the
  Solar System - Dynamics and Evolution, Space Physics, and Spacetime
  Structure.} Kluwer, Dordrecht

\bibitem[{{Brumberg}(1991)}]{1991ercm.book.....B}
{Brumberg} V.~A., 1991, {Essential Relativistic Celestial Mechanics}. Adam
  Hilger, Bristol

\bibitem[{{Cai} {et~al}\mbox{.}(2016){Cai}, {Capozziello}, {De Laurentis}, \&
  {Saridakis}}]{2016RPPh...79j6901C}
{Cai} Y.-F., {Capozziello} S., {De Laurentis} M., {Saridakis} E.~N., 2016, Rep.
  Prog. Phys., 79, 106901

\bibitem[{{Capozziello} \& {de Laurentis}(2011)}]{2011PhR...509..167C}
{Capozziello} S., {de Laurentis} M., 2011, Phys. Rep., 509, 167

\bibitem[{{Capozziello} \& {De Laurentis}(2012)}]{2012AnP...524..545C}
{Capozziello} S., {De Laurentis} M., 2012, Ann. Phys.-Berlin, 524, 545

\bibitem[{{Capozziello}, {de Laurentis} \& {Luongo}(2015){Capozziello}, {de
  Laurentis}, \& {Luongo}}]{2015IJMPD..2441002C}
{Capozziello} S., {de Laurentis} M., {Luongo} O., 2015, Int. J. Mod. Phys. D,
  24, 1541002

\bibitem[{{Capozziello} {et~al}\mbox{.}(2015){Capozziello}, {Harko},
  {Koivisto}, {Lobo}, \& {Olmo}}]{2015Univ....1..199C}
{Capozziello} S., {Harko} T., {Koivisto} T., {Lobo} F., {Olmo} G., 2015,
  Universe, 1, 199

\bibitem[{Cardona \& Tejero(1998)}]{Car98}
Cardona J., Tejero J., 1998, ApJ, 493, 52

\bibitem[{{Carroll}(2001)}]{2001LRR.....4....1C}
{Carroll} S.~M., 2001, Living Rev. Relativ., 4, 1

\bibitem[{{Carroll}(2004)}]{2004sgig.book.....C}
{Carroll} S.~M., 2004, {Spacetime and geometry. An introduction to general
  relativity}. Addison Wesley, San Francisco

\bibitem[{{Carroll}, {Press} \& {Turner}(1992){Carroll}, {Press}, \&
  {Turner}}]{1992ARA&A..30..499C}
{Carroll} S.~M., {Press} W.~H., {Turner} E.~L., 1992, Annu. Rev. Astron. Astr.,
  30, 499

\bibitem[{{Casotto}(1993)}]{1993CeMDA..55..209C}
{Casotto} S., 1993, Celest. Mech. Dyn. Astr., 55, 209

\bibitem[{{Chashchina} \& {Silagadze}(2008)}]{2008PhRvD..77j7502C}
{Chashchina} O.~I., {Silagadze} Z.~K., 2008, Phys. Rev. D, 77, 107502

\bibitem[{{Chennamangalam} \& {Lorimer}(2014)}]{2014MNRAS.440L..86C}
{Chennamangalam} J., {Lorimer} D.~R., 2014, MNRAS, 440, L86

\bibitem[{{Clifton} {et~al}\mbox{.}(2012){Clifton}, {Ferreira}, {Padilla}, \&
  {Skordis}}]{2012PhR...513....1C}
{Clifton} T., {Ferreira} P.~G., {Padilla} A., {Skordis} C., 2012, Phys. Rep.,
  513, 1

\bibitem[{Davis \& Griffen(2010)}]{Davis:2010}
Davis T., Griffen B., 2010, Scholarpedia, 5, 4473, revision \#135530

\bibitem[{{De Felice} \& {Tsujikawa}(2010)}]{2010LRR....13....3D}
{De Felice} A., {Tsujikawa} S., 2010, Living Rev. Relativ., 13, 3

\bibitem[{{De Laurentis} {et~al}\mbox{.}(2017){De Laurentis}, {Younsi},
  {Porth}, {Mizuno}, \& {Rezzolla}}]{2017arXiv171200265D}
{De Laurentis} M., {Younsi} Z., {Porth} O., {Mizuno} Y., {Rezzolla} L., 2017,
  arXiv:1712.00265

\bibitem[{{de Martino}, {De Laurentis} \& {Capozziello}(2015){de Martino}, {De
  Laurentis}, \& {Capozziello}}]{2015Univ....1..123D}
{de Martino} I., {De Laurentis} M., {Capozziello} S., 2015, Universe, 1

\bibitem[{{Debono} \& {Smoot}(2016)}]{2016Univ....2...23D}
{Debono} I., {Smoot} G.~F., 2016, Universe, 2, 23

\bibitem[{{Dunsby} {et~al}\mbox{.}(2010){Dunsby}, {Elizalde}, {Goswami},
  {Odintsov}, \& {Saez-Gomez}}]{2010PhRvD..82b3519D}
{Dunsby} P.~K.~S., {Elizalde} E., {Goswami} R., {Odintsov} S., {Saez-Gomez} D.,
  2010, Phys. Rev. D, 82, 023519

\bibitem[{{Eckart} {et~al}\mbox{.}(2017){Eckart}, {H{\"u}ttemann}, {Kiefer},
  {Britzen}, {Zaja{\v c}ek}, {L{\"a}mmerzahl}, {St{\"o}ckler}, {Valencia-S},
  {Karas}, \& {Garc{\'{\i}}a-Mar{\'{\i}}n}}]{2017FoPh...47..553E}
{Eckart} A. {et~al.}, 2017, Found. Phys., 47, 553

\bibitem[{{Einstein}(1917)}]{1917SPAW.......142E}
{Einstein} A., 1917, Sitzber. Preuss. Akad., 142

\bibitem[{{Espriu}(2014)}]{2014AIPC.1606...86E}
{Espriu} D., 2014, in American Institute of Physics Conference Series, Vol.
  1606, American Institute of Physics Conference Series, pp. 86--98

\bibitem[{{Fienga} {et~al}\mbox{.}(2011){Fienga}, {Laskar}, {Kuchynka},
  {Manche}, {Desvignes}, {Gastineau}, {Cognard}, \&
  {Theureau}}]{2011CeMDA.111..363F}
{Fienga} A., {Laskar} J., {Kuchynka} P., {Manche} H., {Desvignes} G.,
  {Gastineau} M., {Cognard} I., {Theureau} G., 2011, Celestial Mechanics and
  Dynamical Astronomy, 111, 363

\bibitem[{{Gillessen} {et~al}\mbox{.}(2017){Gillessen}, {Plewa}, {Eisenhauer},
  {Sari}, {Waisberg}, {Habibi}, {Pfuhl}, {George}, {Dexter}, {von Fellenberg},
  {Ott}, \& {Genzel}}]{2017ApJ...837...30G}
{Gillessen} S. {et~al.}, 2017, ApJ, 837, 30

\bibitem[{{Goddi} {et~al}\mbox{.}(2017){Goddi}, {Falcke}, {Kramer}, {Rezzolla},
  {Brinkerink}, {Bronzwaer}, {Davelaar}, {Deane}, {de Laurentis}, {Desvignes},
  {Eatough}, {Eisenhauer}, {Fraga-Encinas}, {Fromm}, {Gillessen}, {Grenzebach},
  {Issaoun}, {Jan{\ss}en}, {Konoplya}, {Krichbaum}, {Laing}, {Liu}, {Lu},
  {Mizuno}, {Moscibrodzka}, {M{\"u}ller}, {Olivares}, {Pfuhl}, {Porth},
  {Roelofs}, {Ros}, {Schuster}, {Tilanus}, {Torne}, {van Bemmel}, {van
  Langevelde}, {Wex}, {Younsi}, \& {Zhidenko}}]{2017IJMPD..2630001G}
{Goddi} C. {et~al.}, 2017, Int. J. Mod. Phys. D, 26, 1730001

\bibitem[{{Hees} {et~al}\mbox{.}(2017){Hees}, {Do}, {Ghez}, {Martinez}, {Naoz},
  {Becklin}, {Boehle}, {Chappell}, {Chu}, {Dehghanfar}, {Kosmo}, {Lu},
  {Matthews}, {Morris}, {Sakai}, {Sch{\"o}del}, \&
  {Witzel}}]{2017PhRvL.118u1101H}
{Hees} A. {et~al.}, 2017, Phys. Rev. Lett., 118, 211101

\bibitem[{{Hees} {et~al}\mbox{.}(2014){Hees}, {Folkner}, {Jacobson}, \&
  {Park}}]{2014PhRvD..89j2002H}
{Hees} A., {Folkner} W.~M., {Jacobson} R.~A., {Park} R.~S., 2014, Phys. Rev. D,
  89, 102002

\bibitem[{{Hees} {et~al}\mbox{.}(2012){Hees}, {Lamine}, {Reynaud}, {Jaekel},
  {Le Poncin-Lafitte}, {Lainey}, {F{\"u}zfa}, {Courty}, {Dehant}, \&
  {Wolf}}]{2012CQGra..29w5027H}
{Hees} A. {et~al.}, 2012, Classical Quant. Grav., 29, 235027

\bibitem[{Iorio(2006)}]{Iorio06}
Iorio L., 2006, Int. J. Mod. Phys. D, 15, 473–475

\bibitem[{Iorio(2008)}]{Iorio08}
Iorio L., 2008, Adv. Astron., 2008, 268647

\bibitem[{{Iorio}(2017)}]{2017EPJC...77..439I}
{Iorio} L., 2017, Eur. Phys. J. C, 77, 439

\bibitem[{{Iorio}, {Radicella} \& {Ruggiero}(2015){Iorio}, {Radicella}, \&
  {Ruggiero}}]{2015JCAP...08..021I}
{Iorio} L., {Radicella} N., {Ruggiero} M.~L., 2015, J. Cosmol. Astropart.
  Phys., 8, 021

\bibitem[{{Iorio} {et~al}\mbox{.}(2016){Iorio}, {Ruggiero}, {Radicella}, \&
  {Saridakis}}]{2016PDU....13..111I}
{Iorio} L., {Ruggiero} M.~L., {Radicella} N., {Saridakis} E.~N., 2016, Phys.
  Dark Univ., 13, 111

\bibitem[{{Iorio} \& {Saridakis}(2012)}]{2012MNRAS.427.1555I}
{Iorio} L., {Saridakis} E.~N., 2012, MNRAS, 427, 1555

\bibitem[{Islam(1983)}]{Islam83}
Islam J., 1983, Phys. Lett. A, 97, 239–241

\bibitem[{Jetzer \& Sereno(2006)}]{Jetz06}
Jetzer P., Sereno M., 2006, Phys. Rev. D, 73, 044015

\bibitem[{Kagramanova, Kunz \& L\"{a}mmerzahl(2006)Kagramanova, Kunz, \&
  L\"{a}mmerzahl}]{Kagra06}
Kagramanova V., Kunz J., L\"{a}mmerzahl C., 2006, Phys. Lett. B, 634, 465–470

\bibitem[{Kerr, Hauck \& Mashhoon(2003)Kerr, Hauck, \& Mashhoon}]{Kerr03}
Kerr A., Hauck J., Mashhoon B., 2003, Classical Quant. Grav., 20, 2727–2736

\bibitem[{{Kottler}(1918)}]{1918AnP...361..401K}
{Kottler} F., 1918, Ann. Phys.-Berlin, 361, 401

\bibitem[{Kraniotis \& Whitehouse(2003)}]{Kran03}
Kraniotis G., Whitehouse S., 2003, Classical Quant. Grav., 20, 4817–4835

\bibitem[{{Lucy}(2014)}]{2014A&A...563A.126L}
{Lucy} L.~B., 2014, A\&A, 563, A126

\bibitem[{{Merritt} {et~al}\mbox{.}(2011){Merritt}, {Alexander}, {Mikkola}, \&
  {Will}}]{2011PhRvD..84d4024M}
{Merritt} D., {Alexander} T., {Mikkola} S., {Will} C.~M., 2011, Phys. Rev. D,
  84, 044024

\bibitem[{{Mielke}(2011)}]{2011PhLB..702..187M}
{Mielke} E.~W., 2011, Phys. Lett. B, 702, 187

\bibitem[{{Milani}, {Nobili} \& {Farinella}(1987){Milani}, {Nobili}, \&
  {Farinella}}]{Nobilibook87}
{Milani} A., {Nobili} A., {Farinella} P., 1987, {Non-gravitational
  perturbations and satellite geodesy}. Adam Hilger, Bristol

\bibitem[{{Nesseris} \& {Perivolaropoulos}(2008)}]{2008PhRvD..77b3504N}
{Nesseris} S., {Perivolaropoulos} L., 2008, Phys. Rev. D, 77, 023504

\bibitem[{{Nojiri} \& {Odintsov}(2007{\natexlab{a}})}]{2006hep.th....1213N}
{Nojiri} S., {Odintsov} S.~D., 2007{\natexlab{a}}, Int. J. Geom. Meth. Mod.
  Phys., 4, 115

\bibitem[{{Nojiri} \& {Odintsov}(2007{\natexlab{b}})}]{2007JPhA...40.6725N}
{Nojiri} S., {Odintsov} S.~D., 2007{\natexlab{b}}, J. Phys. A Math. Gen., 40,
  6725

\bibitem[{{Nojiri} \& {Odintsov}(2011)}]{2011PThPS.190..155N}
{Nojiri} S., {Odintsov} S.~D., 2011, Prog. Theor. Phys. Supp., 190, 155

\bibitem[{{O'Raifeartaigh} {et~al}\mbox{.}(2018){O'Raifeartaigh}, {O'Keeffe},
  {Nahm}, \& {Mitton}}]{2017arXiv171106890O}
{O'Raifeartaigh} C., {O'Keeffe} M., {Nahm} W., {Mitton} S., 2018, Eur. Phys. J.
  H

\bibitem[{{Ovcherenko} \& {Silagadze}(2016)}]{2015arXiv151104829O}
{Ovcherenko} S.~S., {Silagadze} Z.~K., 2016, Ukr. J. Phys., 61, 342

\bibitem[{{Padmanabhan}(2003)}]{2003PhR...380..235P}
{Padmanabhan} T., 2003, Phys. Rep., 380, 235

\bibitem[{{Peebles} \& {Ratra}(2003)}]{2003RvMP...75..559P}
{Peebles} P.~J., {Ratra} B., 2003, Rev. Mod. Phys., 75, 559

\bibitem[{{Perlmutter} {et~al}\mbox{.}(1999){Perlmutter}, {Aldering},
  {Goldhaber}, {Knop}, {Nugent}, {Castro}, {Deustua}, {Fabbro}, {Goobar},
  {Groom}, {Hook}, {Kim}, {Kim}, {Lee}, {Nunes}, {Pain}, {Pennypacker},
  {Quimby}, {Lidman}, {Ellis}, {Irwin}, {McMahon}, {Ruiz-Lapuente}, {Walton},
  {Schaefer}, {Boyle}, {Filippenko}, {Matheson}, {Fruchter}, {Panagia},
  {Newberg}, {Couch}, \& {Project}}]{1999ApJ...517..565P}
{Perlmutter} S. {et~al.}, 1999, ApJ, 517, 565

\bibitem[{{Pfahl} \& {Loeb}(2004)}]{2004ApJ...615..253P}
{Pfahl} E., {Loeb} A., 2004, ApJ, 615, 253

\bibitem[{{Planck Collaboration} {et~al}\mbox{.}(2016){Planck Collaboration},
  {Ade}, {Aghanim}, {Arnaud}, {Ashdown}, {Aumont}, {Baccigalupi}, {Banday},
  {Barreiro}, {Bartlett}, \& et~al.}]{2016A&A...594A..13P}
{Planck Collaboration} {et~al.}, 2016, A\&A, 594, A13

\bibitem[{{Psaltis}, {Wex} \& {Kramer}(2016){Psaltis}, {Wex}, \&
  {Kramer}}]{2016ApJ...818..121P}
{Psaltis} D., {Wex} N., {Kramer} M., 2016, ApJ, 818, 121

\bibitem[{{Rajwade}, {Lorimer} \& {Anderson}(2017){Rajwade}, {Lorimer}, \&
  {Anderson}}]{2017MNRAS.471..730R}
{Rajwade} K.~M., {Lorimer} D.~R., {Anderson} L.~D., 2017, MNRAS, 471, 730

\bibitem[{{Riess} {et~al}\mbox{.}(1998){Riess}, {Filippenko}, {Challis},
  {Clocchiatti}, {Diercks}, {Garnavich}, {Gilliland}, {Hogan}, {Jha},
  {Kirshner}, {Leibundgut}, {Phillips}, {Reiss}, {Schmidt}, {Schommer},
  {Smith}, {Spyromilio}, {Stubbs}, {Suntzeff}, \&
  {Tonry}}]{1998AJ....116.1009R}
{Riess} A.~G. {et~al.}, 1998, AJ, 116, 1009

\bibitem[{{Rindler}(2001)}]{2001rsgc.book.....R}
{Rindler} W., 2001, {Relativity: special, general, and cosmological}. Oxford
  University Press, Oxford, UK

\bibitem[{{Ruggiero} \& {Iorio}(2007)}]{2007JCAP...01..010R}
{Ruggiero} M.~L., {Iorio} L., 2007, J. Cosmol. Astropart. Phys., 1, 010

\bibitem[{{Sadeghian} \& {Will}(2011)}]{2011CQGra..28v5029S}
{Sadeghian} L., {Will} C.~M., 2011, Classical Quant. Grav., 28, 225029

\bibitem[{{Sch{\"u}cker} \& {Zaimen}(2008)}]{2008A&A...484..103S}
{Sch{\"u}cker} T., {Zaimen} N., 2008, A\&A, 484, 103

\bibitem[{{Seeliger}(1895)}]{1895AN....137..129S}
{Seeliger} H., 1895, Astron. Nachr., 137, 129

\bibitem[{Sereno \& Jetzer(2006)}]{Ser06}
Sereno M., Jetzer P., 2006, Phys. Rev. D, 73, 063004

\bibitem[{Sereno \& Jetzer(2007)}]{Ser07}
Sereno M., Jetzer P., 2007, Phys. Rev. D, 75, 064031

\bibitem[{{Soffel}(1989)}]{1989racm.book.....S}
{Soffel} M.~H., 1989, {Relativity in Astrometry, Celestial Mechanics and
  Geodesy}. Springer-Verlag; Berlin Heidelberg New York

\bibitem[{{Spergel}(2015)}]{2015Sci...347.1100S}
{Spergel} D.~N., 2015, Science, 347, 1100

\bibitem[{{Stuchl{\'{\i}}k} \& {Hled{\'{\i}}k}(1999)}]{1999PhRvD..60d4006S}
{Stuchl{\'{\i}}k} Z., {Hled{\'{\i}}k} S., 1999, Phys. Rev. D, 60, 044006

\bibitem[{{Weinberg}(1989)}]{1989RvMP...61....1W}
{Weinberg} S., 1989, Rev. Mod. Phys., 61, 1

\bibitem[{{Xie} \& {Deng}(2013)}]{2013MNRAS.433.3584X}
{Xie} Y., {Deng} X.-M., 2013, MNRAS, 433, 3584

\bibitem[{{Zhang} \& {Iorio}(2017)}]{2017ApJ...834..198Z}
{Zhang} F., {Iorio} L., 2017, ApJ, 834, 198

\bibitem[{{Zhang}, {Lu} \& {Yu}(2014){Zhang}, {Lu}, \&
  {Yu}}]{2014ApJ...784..106Z}
{Zhang} F., {Lu} Y., {Yu} Q., 2014, ApJ, 784, 106

\bibitem[{{Zhang} \& {Saha}(2017)}]{2017arXiv170908341Z}
{Zhang} F., {Saha} P., 2017, ApJ, X, Y

\end{thebibliography}


\end{document}